\documentclass{article}
\usepackage{spconf,amsmath,graphicx}
\usepackage{booktabs}

\title{A MULTI-SPEAKER MULTI-LINGUAL VOICE CLONING SYSTEM BASED ON VITS2 FOR LIMMITS 2024 CHALLENGE}
\name{Xiaopeng Wang$^{1}$, Yi Lu$^{1}$, Xin Qi$^{1}$, Zhiyong Wang$^{1}$, Yuankun Xie$^{2}$, Shuchen Shi$^{3}$, Ruibo Fu$^{4,*}\thanks{$^{*}$ Corresponding author.}$}
\address{
  $^1$School of Artificial Intelligence, University of Chinese Academy of Sciences \\
  $^2$School of Information and Communication Engineering, Communication University of China \\
  $^3$Shanghai Polytechnic University 
  $^4$Institute of Automation, Chinese Academy of Sciences \\
    \small{
        \ wangxiaopeng22@mails.ucas.ac.cn, ruibo.fu@nlpr.ia.ac.cn.} \\ 
}

\begin{document}
%
\maketitle
\begin{abstract}
This paper presents the development of a speech synthesis system for the LIMMITS'24 Challenge, focusing primarily on Track 2. The objective of the challenge is to establish a multi-speaker, multi-lingual Indic Text-to-Speech system with voice cloning capabilities, covering seven Indian languages with both male and female speakers. The system was trained using challenge data and fine-tuned for few-shot voice cloning on target speakers. Evaluation included both mono-lingual and cross-lingual synthesis across all seven languages, with subjective tests assessing naturalness and speaker similarity. Our system uses the VITS2 architecture, augmented with a multi-lingual ID and a BERT model to enhance contextual language comprehension. In Track 1, where no additional data usage was permitted, our model achieved a Speaker Similarity score of 4.02. In Track 2, which allowed the use of extra data, it attained a Speaker Similarity score of 4.17.
\end{abstract}
\begin{keywords}
Multi-speaker, Multi-lingual, Text to Speech,  VITS2, LIMMITS'24 Challenge
\end{keywords}
\section{Introduction}
\label{sec:intro}
The LIMMITS'24 Challenge\cite{limmit24}, part of ICASSP 2024, represents a milestone in the advancement of multi-speaker, multi-lingual Text-to-Speech (TTS) models\cite{fu2020focusing}. This year's challenge expands beyond last year's focus on Marathi, Hindi, and Telugu to include Bengali, Chhattisgarhi, English, and Kannada, resulting in a corpus covering seven languages and fourteen speakers. Participants in the challenge can explore TTS voice cloning using a multilingual base model with fourteen speakers. The challenge's complexity is heightened by allowing training with additional multi-speaker corpora like VCTK and LibriTTS, and by introducing zero-shot voice cloning scenarios. A total of 560 hours of high-quality TTS data in seven Indian languages, including the less-resourced Chhattisgarhi, is provided to support participants' endeavors. Rigorous evaluation encompasses both mono and cross-lingual synthesis, evaluated through subjective tests on naturalness and speaker similarity.

In Track 1, with restrictions on additional training data, we implemented VITS2\cite{VITS2}, a single-stage, multi-speaker TTS model. We made several modifications to the model, including integrating a language condition into the text encoder to accommodate the competition's multi-lingual input requirements. This adaptation resulted in a Mean Opinion Score (MOS) of 3.04 for Naturalness and a Speaker Similarity Score of 4.02 in Track 1.

In Track 2, which allowed the use of additional training data, we continued to utilize the VITS2 architecture. Unlike Track 1, we incorporated IndicBERT\cite{indicBERTv2} before the text encoder to enhance contextual understanding and improve prosodic performance. Regarding training data, we combined the competition's training data with the VCTK dataset (excluding target speakers) to enhance the model's accuracy and naturalness across a broader data domain. This approach yielded an MOS of 3.12 for Naturalness and a Speaker Similarity Score of 4.17 in Track 2.
\section{SYSTEM DESCRIPTION}
\label{sec:format}

\subsection{Text preprocessing}
\label{ssec:subhead}

Our system utilizes phonemes as input and employs the espeak\cite{phonemizer} tool to convert Bengali, Kannada, Telugu, Hindi, and Marathi into the International Phonetic Alphabet (IPA). Since espeak does not support Chhattisgarhi, we implemented an alternative method, converting Chhattisgarhi into the IPA for Hindi, a closely related language. Although this approach is not ideal, it provided a feasible solution for our system to handle multiple languages, including Chhattisgarhi, in the absence of direct support.

\subsection{Model Architecture}
\label{ssec:subhead}
\begin{figure}
    \centering
    \includegraphics[width=1\linewidth]{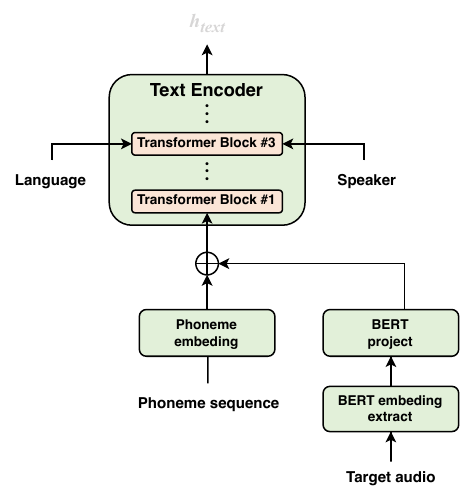}
    \caption{multilingual multi-speaker text encoder with BERT}
    \label{fig:enter-label}
\end{figure}

Our system adopts an architecture based on VITS2, incorporating several modifications to enhance its performance, as depicted in Figure 1. Phoneme sequences are transformed into numerical representations through an embedding process. These embedding vectors are processed by a series of Transformer Blocks within the Text Encoder, utilizing self-attention mechanisms to extract features. We employ IndicBERT, pre-trained on 23 major Indian languages and English, to provide contextual semantic information and enhance the model's expressiveness. The word-level features extracted by BERT are replicated to match the length of phonemes and fused together before being input into the text encoder. Within the text encoder, we merge phoneme information enriched with contextual data with language IDs and speaker IDs to capture diverse language attributes and speaker identities. Ultimately, the synthesized speech is output as target audio.

\subsection{Model Training}
\subsubsection{Pre-training}

We trained the base model using 8 NVIDIA RTX 4090 GPUs with the AdamW optimizer. The batch size was set to 16 and the learning rate to 2e-4. The base model was trained with TTS training data from 7 languages for 136K iterations.
\subsubsection{Fine-tuning}

To achieve few-shot voice cloning on target speakers, we fine-tuned the base model with the few-shot data using a single NVIDIA RTX 4090 GPU. The batch size was set to 16 and the learning rate to 2e-4. After 50K iterations of fine-tuning, the model was capable of synthesizing speech in the voices of 9 target speakers in cross-lingual scenarios.
\section{RESULTS}
The evaluation metrics of the LIMMITS'24 Challenge include naturalness and speaker similarity MOS, covering all 7 Indian languages. We participated in both Track 1 and Track 2, and the scores for both metrics are shown in Table 1, with bold font indicating the first place in that track. We achieved second place in speaker similarity in Track 1 and first place in speaker similarity in Track 2, significantly outperforming the other teams.
\begin{table}[htbp]
    \centering
    \small
    \begin{tabular}{ccccc}
        \toprule
        & \multicolumn{2}{c}{Naturalness} & \multicolumn{2}{c}{Speaker similarity}\\
        \cmidrule(lr){2-3} \cmidrule(lr){4-5}
        & Score (avg) & Score (std) & Score (avg) & Score (std) \\
        \midrule
        Track 1 & 3.04 & 1.15 & 4.02 & 1.178 \\
        Track 2 & 3.12 & 1.09 & \textbf{4.17} & \textbf{1.1184} \\
        \bottomrule
    \end{tabular}
    \caption{MOS results in LIMMITS'24 Challenge}
    \label{tab:my_label}
\end{table}

\section{CONCLUSION}
Our voice synthesis system for the LIMMITS Challenge 2024 demonstrated strong performance in both Track 1 and Track 2, achieving high speaker similarity scores of 4.02 and 4.17, respectively. The use of the VITS2 architecture, enhanced with multi-lingual ID and a BERT model, effectively supported multi-speaker, multi-lingual Indic TTS with robust voice cloning capabilities.
\label{sec:refs}

\bibliographystyle{IEEEbib}
\bibliography{main}

\begin{thebibliography}{1}

\bibitem{limmit24}
Abhayjeet Singh, Amala Nagiredd, Deekshitha G, Jesuraja Bandekar, Roopa R,
  Sandhya Badiger, Sathvik Udupa, Prasanta~Kumar Ghosh, Hema~A Murthy, Pranaw
  Kumar, Keiichi Tokuda, Mark Hasegawa-Johnson, and Philipp Olbrich,
\newblock ``{LIMMITS'24: Multi-speaker, Multi-lingual Indic TTS with voice
  cloning},''
\newblock {\em submitted to ICASSP 2024}, 2024,
\newblock \url{https://sites.google.com/view/limmits24/}.

\bibitem{fu2020focusing}
Ruibo Fu, Jianhua Tao, Zhengqi Wen, Jiangyan Yi, and Tao Wang,
\newblock ``Focusing on attention: prosody transfer and adaptative optimization
  strategy for multi-speaker end-to-end speech synthesis,''
\newblock in {\em ICASSP 2020-2020 IEEE International Conference on Acoustics,
  Speech and Signal Processing (ICASSP)}. IEEE, 2020, pp. 6709--6713.

\bibitem{VITS2}
Jungil Kong, Jihoon Park, Beomjeong Kim, Jeongmin Kim, Dohee Kong, and Sangjin
  Kim,
\newblock ``{VITS2: Improving Quality and Efficiency of Single-Stage
  Text-to-Speech with Adversarial Learning and Architecture Design},''
\newblock in {\em Proc. INTERSPEECH 2023}, 2023, pp. 4374--4378.

\bibitem{indicBERTv2}
Sumanth Doddapaneni, Rahul Aralikatte, Gowtham Ramesh, Shreyansh Goyal,
  Mitesh~M. Khapra, Anoop Kunchukuttan, and Pratyush Kumar,
\newblock ``Towards leaving no indic language behind: Building monolingual
  corpora, benchmark and models for indic languages,''
\newblock {\em ACL}, vol. abs/2212.05409, 2023.

\bibitem{phonemizer}
Mathieu Bernard and Hadrien Titeux,
\newblock ``Phonemizer: Text to phones transcription for multiple languages in
  python,''
\newblock {\em Journal of Open Source Software}, vol. 6, no. 68, pp. 3958,
  2021.

\end{thebibliography}

\end{document}